\documentclass[12pt,english]{article}
\usepackage{latexsym}
\usepackage{epsfig}
\usepackage{a4}
\usepackage{amsfonts}
\renewcommand{\theequation}{\thesection.\arabic{equation}}
\newcounter{subequation}[equation]
\makeatletter

\expandafter\let\expandafter\reset@font\csname reset@font\endcsname

\def\subeqnarray{\arraycolsep1pt
    \def\@eqnnum\stepcounter##1{\stepcounter{subequation}%
	{\reset@font\rm(\theequation\alph{subequation})}}
\jot5mm     \eqnarray}

\makeatother

\def\be{\begin{equation}}
\def\ee{\end{equation}}
\def\bea{\begin{eqnarray}}
\def\eea{\end{eqnarray}}

\def\ket#1{\left|#1\right>}
\def\bra#1{\left<#1\right|}

\def\one#1{#1^{\raise5pt\hbox{$\scriptstyle\!\!\!\!1$}}\,{}}
\def\two#1{#1^{\raise5pt\hbox{$\scriptstyle\!\!\!\!2$}}\,{}}

\def\tilde{\widetilde}
\makeatletter
\def\binrel@#1{\begingroup
  \setboxz@h{\thinmuskip0mu
    \medmuskip\m@ne mu\thickmuskip\@ne mu
    \setbox\tw@\hbox{$#1\m@th$}\kern-\wd\tw@
    ${}#1{}\m@th$}%
  \edef\@tempa{\endgroup\let\noexpand\binrel@@
    \ifdim\wdz@<\z@ \mathbin
    \else\ifdim\wdz@>\z@ \mathrel
    \else \relax\fi\fi}%
  \@tempa
}
\let\binrel@@\relax
\def\overset#1#2{\binrel@{#2}%
  \binrel@@{\mathop{\kern\z@#2}\limits^{#1}}}
\def\underset#1#2{\binrel@{#2}%
  \binrel@@{\mathop{\kern\z@#2}\limits_{#1}}}
\makeatother
\newfont{\bbd}{msbm10 scaled\magstep1}

\def\om{\omega}

\newcommand{\p}{^{\prime}}
\begin{document}
\hfill NTZ 10/2002

{\begin{center}
{\LARGE {Electroproduction of vector mesons - \\ 
factorization of end-point contributions} } \\ [8mm]
{\large  A. Ivanov\footnote{e-mail: 
Alexander.Ivanov@itp.uni-leipzig.de} and 
R. Kirschner \\ [3mm] } 

 Naturwissenschaftlich-Theoretisches Zentrum und 
Institut f\"{u}r Theoretische Physik, Universit\"{a}t Leipzig, \\
Augustusplatz 10, D-04109 Leipzig, Germany

\end{center} 
} 

\vspace{3cm} 
\noindent 
{\bf Abstract}

\noindent
The end-point contributions in the quark longitudinal momentum fraction of
the virtual photon ($\gamma^*$) to vector meson ($V$) impact factor
to the diffractive electroproduction amplitude can be factorized in terms of
a generalized parton evolution of the target parton distribution.
The result is used to model the helicity amplitudes $\gamma^* p \rightarrow
V p$ in terms of small $x$ generalized parton distributions.    

\newpage



\section{Introduction}
\setcounter{equation}{0}

The experimental analysis of diffractive vector meson production by
virtual or quasi-real photons ($\gamma p \rightarrow V p^* $) at HERA 
\cite{c,Abramowicz:1998ii,Breitweg:1999jy,Breitweg:1999fm,Chekanov:2002rm,
Chekanov:2002xi,Adloff:2000nx,Adloff:2000vm,Adloff:1999kg,Adloff:2002tb} 
has been accompanied by numerous theoretical and phenomenological
studies, e.g.
\cite{Ryskin:1992ui,Brodsky:1994kf,Frankfurt:1995jw,Frankfurt:1997fj,
Ginzburg:1996vq,Nemchik:1996pp,Nemchik:1996cw,Martin:1996bp,Royen:1998zk,
Martin:1999wb,Ivanov:1998gk,Kuraev:1998ht,Kirschner:qq}. 
 The main questions under discussion are
the typical ones for semi-hard processes: To what extend perturbative
QCD applies? What can we learn about non-perturbative hadronic interactions?

The factorization proof \cite{Collins:1997hv} gives a first answer. 
In the 
helicity amplitudes with $\lambda_i (\gamma) =0$ factorization
holds and the contribution of a small-size $q \bar q$ dipole coupled
by two gluons to the exchanged pomeron dominates by power counting
 the contributions with additional soft exchanges. For the remaining
helicity amplitudes the power counting does not result in the dominance
of the short distance contribution. This is different from the related
deep virtual Compton amplitude (DVCS, $ \gamma^* p \rightarrow \gamma^* p $)
where the short distance contribution dominates in all helicity amplitudes
 \cite{Collins:1998be}.

In some presentations the situation is commented by calling the dominating
short distance contribution in the helicity amplitude $\lambda_i = \lambda_f = 0$
the leading twist one and by saying that short distance factorization
must not be expected for the other helicity amplitudes because of
being of non-leading twist. 

Owing to the DVCS case the latter argument
looks not convincing. Moreover, the amplitudes with helicities 
different from
$\lambda_i = \lambda_f =0 $ are accessible to experiment, e.g. in
the ratio of the cross sections of longitudinal to transverse virtual
photons and in the decay-angular distributions parametrized by the
Schilling-Wolf ratios \cite{Schilling:1973ag,c}.
If short distance factorization
holds the amplitude can be expressed in terms of generalized parton
distributions of the target and  predictions for $\sigma_L / \sigma_T $ 
\cite{Martin:1996bp,Nemchik:1996pp,Nemchik:1996cw,Royen:1998zk,Martin:1999wb}
and the Schilling-Wolf ratios \cite{Ivanov:1998gk,Kuraev:1998ht}
 have been obtained. In some of the approaches
 the amplitudes have been treated in analogy to the one with
$\lambda_i = \lambda_f =0 $, i.e. with the $q \bar q$ dipole interacting
via two gluon exchange coupled to the generalized gluon distribution.
However, unlike the $\lambda_i = \lambda_f = 0 $ case, in the leading twist
contribution to the other helicity amplitudes singularities 
in the momentum fraction
of the quark in the dipole appear. They are related to a large transverse
size of the dipole and seem to signal the factorization breakdown
expected from the power counting ana\-lysis \cite{Collins:1997hv}. 

There are controversial
opinions about the appropriate treatment of those end-point singularities:
introducing physically motivated cut-offs, choosing damping meson
wave functions or including damping quark formfactors, or relating the
argument of the parton distributions to the increasing dipole size.
In any case, the partial success of these perturbative approaches
seem to result in a modification of the first answer to the above
main questions as given by the factorization proof: The factorization
breaking effects cannot be large for all helicity amplitudes. 

The
connection of the end-point singularities to the factorization breaking
is referred to frequently. In the study \cite{Mankiewicz:1999tt} using the
operator product language these end-point singularities appear as
the obstacle to factorization.

In the present paper we end up with the opposite conclusion: The end-point
contributions are factorizable; the ones appearing in the quark dipole
interacting by two gluon exchange are factorized by identifying
them as a leading $\ln Q^2$ Bjorken evolution term of the two gluon
($gg$ )to a quark - anti quark   ($q \bar q$ ) exchange. 
Our discussion relies on a specific model
of the meson light-cone wave function. It  includes more than the leading
twist contribution in terms of distribution amplitudes. Although finally
only a small part of the specific information encoded in this wave
function enters the results on the large $Q^2$ asymptotics the
inclusion  in the wave function and resummation of 
a geometric series of higher twist terms
$\sim ({m_V^2 \over Q^2 z \bar z})^n$ is the essential point for understanding
the physical meaning of the end-point contributions. 

In section 2 we start
with the impact factor representation of the diffractive amplitude
and specify to the contribution of a scattering $q \bar q$ dipole
coupled to the pomeron exchange by two gluons. In section 3 the $\gamma^* V$
impact factors are introduced by specifying the meson light cone wave
function. The asymptotics for large $Q^2$ is calculated. We compare
with the $\gamma^* \gamma^*$ impact factor and continue this comparison
in the following. Some details of the calculations are given in the
Appendix. In section 4 the logarithmic end-point contributions to
the amplitudes are identified with a leading $\ln Q^2 $ contribution
to the generalized Bjorken evolution. In order to make this main point
clearer this identification is repeated for the 
$\gamma^* p \rightarrow \gamma^* p $
amplitude with $ \lambda_i = \lambda_f =1 $, i.e. for
the well known case related to the structure function $F_2 (x, Q^2)$. 
In section 6 we evaluate
numerically the resulting leading twist terms of the helicity amplitudes,
specifying a the small $x$ parton distribution, and
obtain results for quantities which can be compared with the results
of the experimental analysis: the cross section ratio $\sigma_L / \sigma_T$
and the angular-decay  distribution (Schilling-Wolf) coefficients 
$ r^{\alpha}_{jk}$ in dependence on $Q^2$ and $t$.

\section{Diffractive \protect\( \gamma \protect \){*}V amplitudes}
\setcounter{equation}{0}

A good starting point for analyzing high energy diffractive processes
is the impact factor representation of the corresponding amplitude
written in terms of partial waves,
$$
M^{\lambda _{i}\lambda _{f}}(s,Q,q)=s\int ^{i\infty }_{-i\infty }
\frac{d\omega }{2\pi i}
F^{^{\lambda _{i}\lambda _{f}}}(\omega ,Q,q) 
\left[ \left( \frac{s}{M^{2}(Q,m,q)}\right) ^{\omega }+
\left( \frac{-s}{M^{2}(Q,m,q)}\right) ^{\omega }\right], $$ 
\be
\label{partial_waves}
F^{^{\lambda _{i}\lambda _{f}}}(\omega ,Q,q)=
\int d^{2}\kappa d^{2}\kappa^{\prime }
\Phi ^{^{\lambda _{i}\lambda _{f}}}(\kappa,Q,q) \ 
{\cal G} (\kappa, \kappa^{\prime },q,\omega ) \ 
\Phi ^{P}(\kappa^{\prime },q)
\ee

This is a typical form obtained in perturbative analysis
\cite{ChW1,Lev89},
however it is based on more general arguments relying on impulse 
approximation \cite{Gribov:jg}.
The field representing the exchange interaction acts on the scattering
particles for a short time, much shorter than the time scale of their
binding or self-interaction. The field sees just a 
short-time intermediate fluctuation state. 
The impact factors appear as the transition matrix elements
(\( \gamma ^{*}\rightarrow V \) and \( P\rightarrow P^{*} \) in
our case ) of an operator representing the action of the exchange
field on the intermediate fluctuations selecting some of the fluctuations
according to their interaction strength.

The diffractive exchange $ {\cal G} $ is called Pomeron. In phenomenology
it is often substituted by a Regge pole or, according to a resent
proposal \cite{Donnachie:2001xx}, by two poles, a soft and a hard Pomeron. 
In the
framework of QCD we suppose the Pomeron to consist out of interacting
gluons. In perturbative QCD this idea acquires a definite meaning
in the BFKL scheme \cite{BFKL,Lev89}. 

In the case of hard diffraction the intermediate fluctuations is squeezed
into a narrow space-time region (\( \sim \frac{1}{Q} \)) in the vicinity
of the light cone. Only short-distance modes of Pomeron field can
interact with this fluctuation state. In this situation both the fluctuation
state and the coupling to the Pomeron can be represented by perturbative
QCD. Moreover one finds, that the fluctuations with a small number
of partons dominates, the higher Fock states being suppressed by 
powers of 
\( \frac{1}{Q} \). Also the exchange with a minimal number of exchange
partons dominates not only by small couplings but, more important, by
powers of \( \frac{1}{Q} \). This suppression holds if the exchanged
partons carry large transverse momenta $\kappa$ , 
\( Q^{2}\gg \kappa^{2}\gg m^{2} \),
and one has to make sure that contributions with extra soft exchange quanta 
are absent or can be absorbed into generalized parton distributions.

In the diffractive vector meson production by  virtual photons
the factorization in terms of a \( q\overline{q} \) intermediate
fluctuation coupling to  the pomeron by two gluons has been proven
on a rigourous level in the case where the short distance scale
 is provided by the momentum
squared of the virtual photon with longitudinal 
polarization \cite{Collins:1997hv}.

In the case of factorization via two exchanged partons one can write
\be
\label{2part_fact}
{\cal G} (k,k^{\prime },q,\omega )=
\frac{1}{|k|^{2}|k+q|^{2}}\widetilde{{\cal G} }(k,k^{\prime },q,\omega )
\ee

In the considered case of electro-production we have in eq.
(\ref{partial_waves} )
\( M^{2}(Q^{2},m,q) \)
\( \approx Q^{2} \). We pick up the leading contribution in the \( \omega  \)
integral by writing 
\be
\label{leading_asymp}
M^{\lambda _{i}\lambda _{j}}=\sum _{p}\int d^{2}\kappa
\Phi _{p}(\kappa,q,Q)\frac{1}{|k|^{2}|\kappa +q|^{2}}
G^{\prime }_{p}(x_1,x_2,q,\kappa)
\ee

\( G_{p}^{\prime } \) stands for the unintegrated generalized parton
distribution of the proton (for recent reviews see 
\cite{Radyushkin:2000uy,Belitsky:2001ns,Goeke:2001tz} )
 resulting from 
the convolution of the proton impact factor $\Phi^P$
with the projection of the exchange \( {\cal G}  \) that couples by
two exchange partons $(p)$ in the small Bjorken variable limit, \( x_1 \)
\( = \frac{Q^{2}}{s} \), $x_2 = \frac{m_V^2}{s}$. 
The skewdness \( \xi = \frac{x_1 - x_2}{2} \) is small;  we
have non-vanishing transverse momentum transfer $q $.

As a simplification we replace \( G_{p}^{\prime } \) by the derivative
of the  gluon distribution at small $x$
\be
\label{unitegratedG}
G^{\prime }_{p}=\frac{\partial }{\partial \ln |\kappa|^{2}}
[ G_{p}(x_1,x_2,|q|^2, |\kappa|^{2}) \ T(\kappa^2, Q^2)) ]
\ee
Here $T(\kappa^2, Q^2)$ is the parton Sudakov formfactor;
it is equal to 1 at $\kappa^2 = Q^2$ and small for $\kappa^2 \ll Q^2$.

The relation to the standard notation introduced by Ji is as follows,
\be
G_{p}(x_1,x_2) = H^{\it Ji}_p ( x, \xi), \ \ x= \frac{x_1 + x_2}{2},
\xi = \frac{x_1 - x_2}{2}.
\ee

For the leading contribution in \( Q^{2} \) we expand 
\( \Phi _{p}^{\lambda _{i}\lambda _{f}}(k,q,Q^{2}) \)

\be
\label{leading_impact}
\Phi ^{^{\lambda _{i}\lambda _{f}}}(\kappa,Q,q)=
\left( \frac{\mu ^{2}(m_{V},q)}{Q^{2}}\right) ^{\tau }
\frac{k(k+q)^{*}+k^{*}(k+q)}{Q^{2}}C^{\lambda _{i}\lambda _{f}}+...
\ee

The leading \( \ln Q^{2} \) contribution of the $\kappa $ integral is then
obtained as
\be
\label{leading_ampl}
M^{\lambda _{i}\lambda _{f}}(s,Q,q)=
\sum _{p}\left( \frac{\mu ^{2}(m_{V},q)}{Q^{2}}\right) ^{\tau }
\frac{1}{Q^{2}}C^{\lambda _{i}\lambda _{f}}G_{p}(x_1,x_2,q, Q^{2})
\end{equation}

As an obstacle to factorization in diffractive electroproduction with
transverse polarization, mentioned in the Introduction, 
one encounters large contributions from 
\( q\overline{q} \)
fluctuation states with the longitudinal momentum fraction of one
of the partons ($z$ or $\bar z = 1-z $) small. 
The soft scattering quark then may
couple by soft exchange gluon to the Pomeron without suppression by
1/Q.

We shall investigate the expressions for impact factors with \( q\overline{q} \)
state and a model wave function \( \psi ^{V} \) of the vector meson.
We shall show that the enhanced end-point contribution at large \( Q^{2} \)
\( (s\gg Q^{2}\gg m^{2}_{V}) \) actually arise from $1  \gg  
z, \bar z  \gg  $, \( \frac{|\kappa|^{2}}{Q^{2}} \). In this range the
$z$ integral is approximately logarithmic and this contribution can
be identified as the one of generalized Bjorken evolution 
\cite{DGLAP,ERBL}
of the two exchanged gluons \( gg \) to exchanged quark-antiquark
\( q\overline{q}. \) The latter exchange involve higher twist modes.

In this way we are going to show 
that the soft quark or end point contribution
can be factorized and included in the parton distribution of the diffractive
exchange. The factorization of this contributions goes via a 
\( q\overline{q} \)
exchange instead of \( gg \). The $gg$ coupling to the Pomeron is 
then restricted
to the contribution of the hard (\( z \), \( \bar z=O(1) \) ) scattering
\( q\overline{q} \) dipole.

\section{Impact factors}
\setcounter{equation}{0}

The impact factor with \( q\overline{q} \) intermediate state and
two leading gluon exchange can be written as (compare Fig. 1)
\bea
\label{dipole}
\Phi ^{^{\lambda _{i}\lambda _{f}}}(\kappa_{1},\kappa_{2})= \int 
d^{2}\ell_{1}d^{2}\ell_{2}dz\psi _{i}^{\lambda _{i}}(l_{1},z)
\phi ^{dipole}(\ell_{1},l_{2},\kappa_{1},\kappa_{2}) 
\psi _{f}^{\lambda _{f} *}(\ell_{2}-zq,z)
\cr
\phi ^{dipole}(\ell_{1},\ell_{2},\kappa_{1},\kappa_{2})=\alpha _{s}
[\delta ^{2}(\ell_{2}-\ell_{1})
+\delta ^{2}(\ell_{2}-\ell_{1}+\kappa_{1}+\kappa_{2})- \cr
\delta ^{2}(\ell_{2}-\ell_{1}+\kappa_{1})-
\delta ^{2}(\ell_{2}-\ell_{1}+\kappa_{2})]
\eea

The first argument in the light-cone wave functions $\psi_{i/f}$ is the 
transverse
momentum relative to the momentum direction of corresponding particle.
\( \kappa_{i} \) are the transverse momenta components of the exchange
gluons, \( \kappa_{1}+\kappa_{2}=-q \) 
is the transverse part of the momentum transfer.

\begin{figure}[htbp]
\begin{center}
\vspace{-0.cm}
\epsfig{file=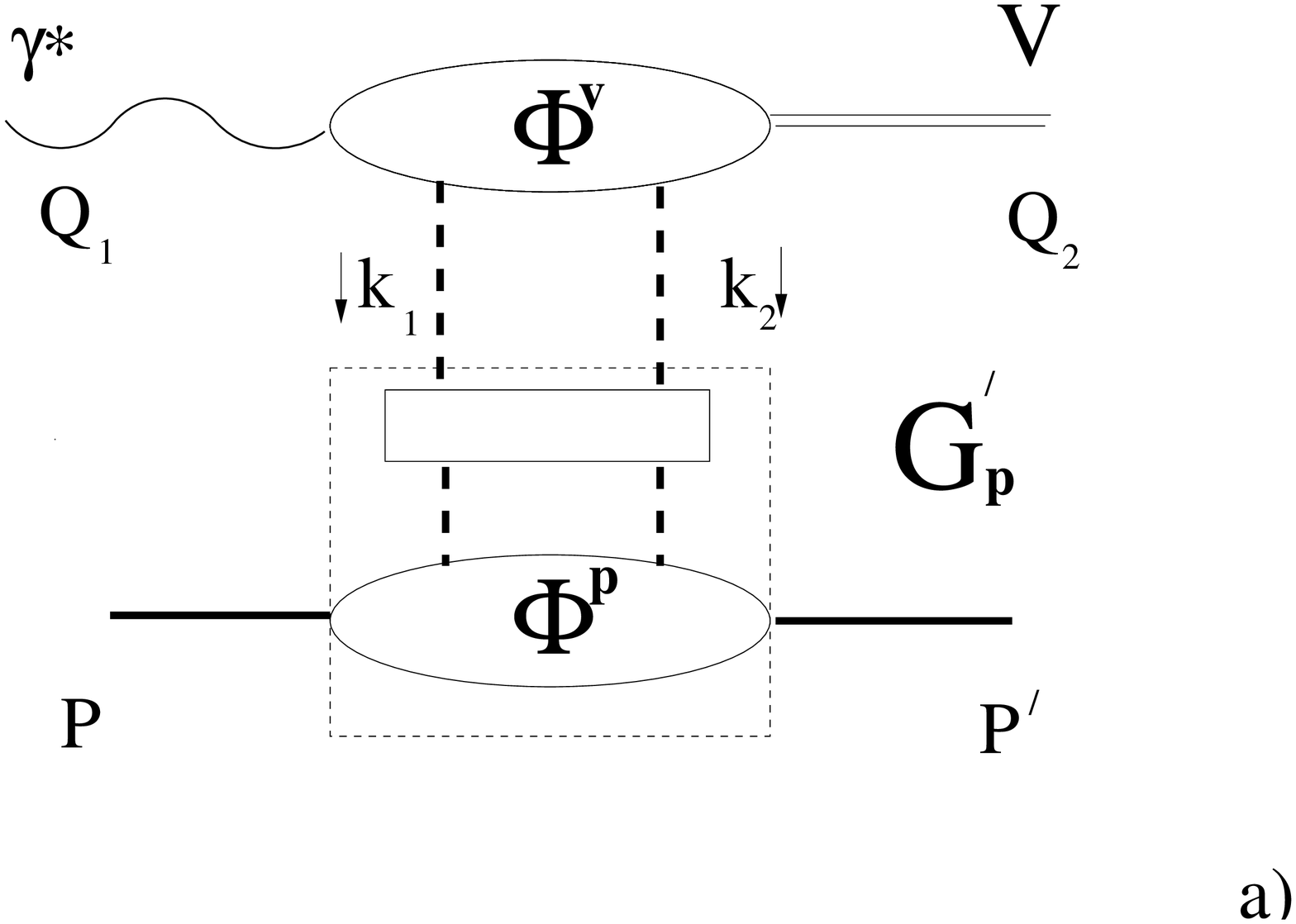,width=10cm}
\epsfig{file=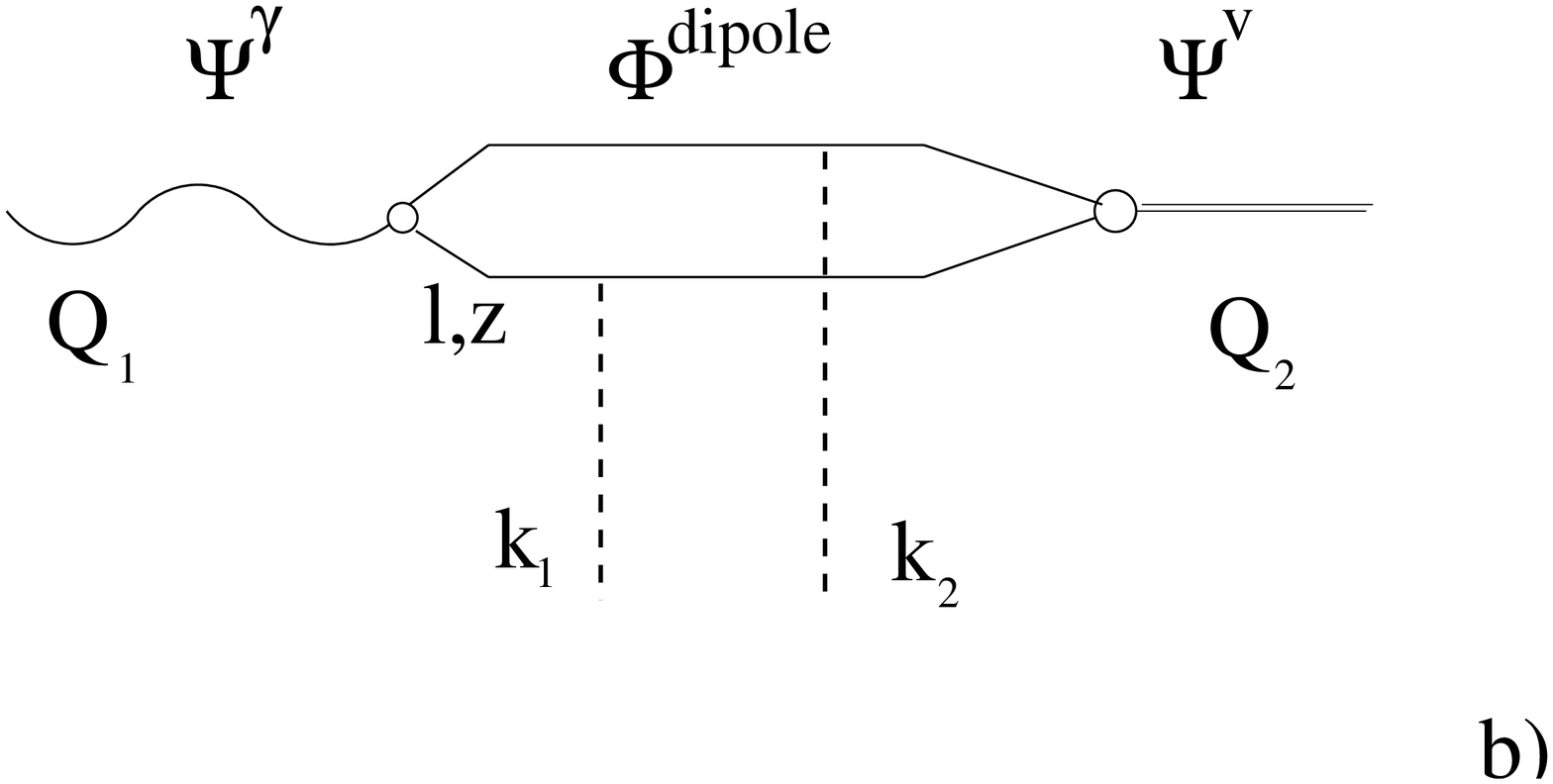,width=10cm}
\end{center}
\caption{a) Impact factor form of the $\gamma^* P \rightarrow V P $
amplitude. The rectangular box indicates the exchange interaction and
the dashed line box the unintegrated generalized gluon distribution.
b) Contribution to the $\gamma^* V $ impact factor.  }
\label{fig:rt}
\end{figure}

The \( \gamma ^{*}\gamma ^{*} \) impact factor can be treated purely
perturbative. It is obtained in the perturbative Regge asymptotics
as the integral over the right hand cut of the discontinuity of the
amplitude \( \gamma ^{*}g\rightarrow \gamma ^{*}g \).
The four terms in \( \phi ^{dipole} \) (\ref{dipole}) correspond
to the four ways to couple the two gluons to \( q\overline{q} \);
one of them is shown in Fig.1b. The momentum variables are defined in
the Sudakov frame
\bea
\label{sudakov} 
Q_{1}=q^{\prime }-x_{1}p,\ \ Q_{2}=q^{\prime }-x_{2}p+q, \cr
\l = zq^{\prime }-\beta_{\ell}  p + \ell, \cr
k_{i}=\alpha _{i}q^{\prime }- \beta _{i}p+\kappa_{i}, \cr
2q^{\prime }p=s 
\end{eqnarray}
The \( \delta  \) functions of the mass shell condition at the right-hand
discontinuity in the subenergy \(- 2k_{1} Q_{1}\approx \beta _{1}s \), 
e.g. for Fig.1b,
 \begin{equation}
\label{mass_shell}
z\delta ((z-\alpha _{1})(\beta _{1}-\beta _{\ell})s-|\ell-\kappa|^{2}-
m^{2}_{q}) \ \bar z\delta (\bar z(\beta _{\ell} - x_{1})s-|\ell|^{2}-m^{2}_{q})
\end{equation}
are used to do the integrals over this subenergy $\beta_1 s$ and also over
the loop momentum Sudakov component $\beta_{\ell}$. The Sudakov components
$\alpha_i$ of the exchanged gluons are neglected in this step.

 The two remaining propagators result in \( \Psi _{i}, \) \( \Psi _{f}, \)
the light-cone wave function of \( \gamma ^{*} \),
\bea
\label{psigamma}
\Psi ^{(\gamma )\lambda }(\ell,z,Q)=
e \frac{V^{\lambda }(\ell,z,Q)}{[Q^{2}+\frac{|\ell|^{2}+m^{2}_{q}}{z\bar{z}}]}
\cr
V^{(0)}=Q,\ V^{(+1)}=\frac{\ell^{*}}{z},\ V^{(-1)}=\frac{\ell}{\overline{z}}
\eea
The sum over fermion chiralities adds the same with 
\( z\leftrightarrow \overline{z} \)
or amounts in the factor 2 in (\ref{dipole})

In the helicity cases \( \lambda _{i}=\lambda _{f}=\pm 1 \) there
is an additional chiral odd term which is included 
by the substitution
\be
\label{chiral_odd}
2V^{+1}(\ell_{1},z)V^{*+1}(\ell_{2},z)=\ell_{1}^* \ell_{2}(\frac{1}{z^{2}}+
\frac{1}{\overline{z}^{2}})+
\frac{m_{q}^{2}}{(z\overline{z})^{2}}.
\ee
As a model for the light cone wave function of the vector meson we
assume
\be
\label{psiV}
\Psi ^{V\lambda }(\ell,z)=
f_{V}\frac{V^{\lambda }(\ell,z,m_{V})}{m^{2}_{V}}
\exp\left[ -\frac{|\ell|^{2}+m^{2}_{q}}{z\overline{z}m_{V}^{2}}\right] 
\ee
The form is motivated by QCD sum rules, it is formally obtained by
Borel transformation of the propagator factor in (\ref{psigamma})
with respect to \( Q^{2} \) and by the substitution of Borel variable
by \( m^{2}_{V} \),   where \( m_{V} \) is of the oder of the meson mass 
 \cite{Balitsky:ns}.
This wave function, being close to the one of \( \gamma ^{*} \),
is a particular realization of the phenomenologically successful concept
of vector dominance.
Actually the explicit form of \( \Psi ^{V} \) involves more information
than necessary for the asymptotic estimate at large \( Q \). The
essential point in changing from \( \Psi ^{\gamma } \) to \( \Psi ^{V} \)
is removing the hard (singular in impact parameter, the Fourier conjugate
to \( \ell \)) component while keeping the helicity structure.

In (\ref{psigamma}, \ref{psiV}) we have kept the quark mass to indicate the
possible extension to the case of heavy quark vector mesons; it will be
neglected in the following.

In the Appendix we consider the impact factors in some detail and
calculate the leading twist contribution, for \( \gamma ^{*}\gamma ^{*} \)
at \( Q_{1}^{2}=x_{1}s \),  \( Q_{2}^{2}=x_{2}s, \) with $x_1, x_2$ small and
\( s\rightarrow \infty  \)
and for \( \gamma ^{*}V \) at $Q^2 \rightarrow \infty $,
for representative cases of polarizations.
For the asymptotic estimate we divide the range in \( z \) 
into \( z_{0}\leq z\leq 1-z_{0}= \bar z_0 \),
\( 0<z<z_{0} \) and \( \bar z_{0}<z<1, \),
\be
\label{division}
\Phi =\Phi _{1}+\Phi _{z_{0}}+\Phi _{\bar z_{0}}.
\ee
The results for \( z=O(1) \) are 
\[
\Phi ^{\lambda _{1}\lambda _{f}}_{1}(q,k)=
\int ^{1-z_{0}}_{z_{0}}\varphi ^{\lambda _{i}\lambda _{f}}_{4}(q,k)
z\overline{z}dz, \]
\[
z\overline{z}\varphi ^{00}_{4}(q,k)=C_{1}^{00}\frac{f^{(2)}(k,q)}
{\widetilde{Q}^{2}}+\]
\[
C^{00}_{2}\left\{ \frac{|f^{(2,**)}|^{2}}{\widetilde{Q}^{4}}\
frac{1}{z\overline{z}}-(4-\frac{1}
{z\overline{z}})\left[ |q|^{2}f^{(2)}(k,q)+
\frac{1}{2}q^{*2}f^{(2,**)}(k,q)+\frac{1}{2}q^{2}f^{(2,**)}(k,q)\right] 
\right\}, \]
 \begin{equation}
\label{main}
z\overline{z}\varphi ^{01}_{4}(q,k)=C^{01}
\frac{f^{(3)}(k,q)}{\widetilde{Q}^{3}}(4-\frac{1}{z\overline{z}}),
\end{equation}
\[
z\overline{z}\varphi ^{10}_{4}(q,k)=C^{10}
\frac{f^{*(3)}(k,q)}{\widetilde{Q}^{3}}(4-\frac{1}{z\overline{z}}),\]

\( z\overline{z}\varphi ^{1-1}_{4}(q,k)=C_{1}^{1-1}
{f^{(2)**}(k,q) \over \widetilde{Q}^{2}}+ \)
\[
C_{2}^{1-1}\frac{1}{\widetilde{Q}^{4}}
\left\{ \frac{1}{2}f^{(2)}(k,q)f^{(2)**}(k,q)
\frac{1}{z\overline{z}}+\left[ q^{*2}f^{(2)}(k,q)-|q|^{2}
f^{(2**)}(k,q)\right] (\frac{1}{z\overline{z}}-3)\right\}, \]
\[
z\overline{z}\varphi ^{11}_{4}(q,k)=C^{11}
\frac{f^{(2)}(k,q)}{\widetilde{Q}^{2}}(\frac{1}{z\overline{z}}-2). \]

We have introduced the abbreviations
\[
f^{(2)}(k,q)=k(k+q)^{*}+k^{*}(k+q), \ f^{(2)**}(k,q)=2k^{*}(k+q)^{*},\]
\begin{equation}
\label{f-functions}
f^{(3)}(k,q)=q^{*}f^{(2)}(k,q)+\frac{1}{2}qf^{(2)**}.
\end{equation}

In the case $\gamma^*V$ the large scale $\tilde Q^2$ is just $Q^2$.
For a smoother extrapolation back into the subasymptotic region
we replace $\tilde Q^2 = Q^2 + m_V^2$. 
In the case $\gamma^* \gamma^* $ the large scale $\tilde Q^2$ is to be
substituted in (\ref{main}) 
by $\tilde Q^2 = s $; actually $s$ enters always multiplied by
$x_1 y + x_2 \bar y$ which we have absorbed into the coefficients 
$C^{\lambda_i\lambda_f} $ in this case.

The coefficients \( C^{\lambda _{i}\lambda _{f}} \) depend on \( x_{1},x_{2} \)
in the case of \( \gamma ^{*}\gamma ^{*}: \)\[
C_{1}^{\gamma 0,0}(x_{1,}x_{2})=-2\sqrt{x_{1}x_{2}} Y(x_{1},x_{2},1,1,2),
\ \ C_{2}^{\gamma 0,0}(x_{1,}x_{2})=2\sqrt{x_{1}x_{2}} Y(x_{1},x_{2},2,2,3)\]
\[
C^{\gamma 0,1}(x_{1,}x_{2})=\sqrt{x_{1}}  Y(x_{1},x_{2},2,1,2)\]
\begin{equation}
\label{photonC}
C^{\gamma 1,0}(x_{1,}x_{2})=-\sqrt{2} Y(x_{1},x_{2},1,2,2)
\end{equation}
\[
C_{1}^{\gamma -1,1}(x_{1,}x_{2})= 2 Y(x_{1},x_{2},1,1,1),
C_{2}^{\gamma -1,1}(x_{1,}x_{2})=-2 Y(x_{1},x_{2},2,2,2)\]
\[
C^{\gamma 1,1}(x_{1,}x_{2})=x_{2} Y(x_{1},x_{2},0,2,2),\]
with the abbreviation
\[
Y(x_{1},x_{2},a,b,c)=\int ^{1}_{0}\
\frac{dyy^{a}\overline{y}^{b}}{(x_{1}y+x_{2}\overline{y})}.\]

In the case \( \gamma ^{*}V \) the coefficients 
\ \( C^{\lambda _{i}\lambda _{f}} \)
depend on \( m_{V} \) and \( Q \)
\[
C_{1}^{V 0,0}=C_{2}^{V 0,0}=-2\frac{m_{V}Q}{Q^{2}+m^{2}_{V}}\]
\[
C^{V 0,1}=\frac{m^{2}_{V}Q}{(m^{2}_{V}+Q^{2})^{3/2}}\]
\begin{equation}
\label{mesonC}
C^{V 1,0}=\frac{m_{V}}{(m^{2}_{V}+Q^{2})^{1/2}}
\end{equation}
\[
C_{1}^{V 1,-1}= - \frac{1}{2} C_{2}^{1,-1}= C^{V 1,1} = 
\frac{m^{2}_{V}}{Q^{2}+m^{2}_{V}}\]

In the asymptotically dominating amplitude $\lambda_i = \lambda_f = 0$ we
have included the next-to leading term in the twist power expansion. This is
done in oder to demonstrate that the higher twist terms are accessible in the
present approach and to test their effect on the numerical estimates.

The leading terms of the end-point contributions from $z={\cal O} (z_0)$
are
\be
\label{end-point}
\Phi ^{\lambda _{i}\lambda _{f}}_{z_{0}} \ 
=
C^{\lambda _{i}\lambda _{f}}_{0}  \frac{f^{(n)}(\kappa,q)}{\tilde Q^{n}}
\ln \frac{\tilde Q^{2}z_{0}}{|\kappa|^{2}}-
\frac{c^{\lambda_i \lambda_f}(k,q)}{\tilde Q^{n}}
\ln \frac{|\kappa|^2}{|q|^2 }  
+ {\cal O } (z_0).
\ee
The contributions from the other end point, $\Phi_{\bar z_0}$, are
obtained by replacing $z_0$ by $\bar z_0$.  
Here \( C_{0}^{\lambda _{i}\lambda _{f}} \) coincide with the coefficients
in front of 
\( \int \frac{1}{z\overline{z}} \) in \( \Phi_1 ^{\lambda _{i} \lambda _{f}} \)
in eq. (\ref{main}) as expected for cancelling the auxiliary \( z_{0} \).

\section{Factorization of end point contributions}
\setcounter{equation}{0}

As the result of the previous section we have separated the end point
contributions. Starting from the impact factor with a reasonable meson
wave function we have identified the end point contributions for the
kinematics \( Q^{2}\gg |k|^{2}\gg  \)\( m^{2}\sim |q|^{2} \) as
being proportional to
\be
\label{end-point_z}
\left (\int ^{z_0}_{\frac{|\kappa|^2}{Q^2} } dz \ + \ 
\int^{1- {|\kappa|^2 \over Q^2} }_{\bar z_0} dz \right )
    \frac{1}{z\overline{z}}.
\ee
We observe that there is no divergence at 
\( z,\overline{z} \)\( \rightarrow 0 \).
Spurious divergences appear only in the extrapolation of the twist
expansion done for \( z=O(1) \) (\ref{main}) to 
\( z,\overline{z} \)\( \rightarrow 0 \).
As we see, the blind extrapolation would neglect terms with powers 
of \( \frac{\kappa^{2}}{z\overline{z}Q^{2}} \).
The  model vector meson wave function  just specifies  
how the sum of these higher twist 
terms removes the end point divergence and leads to (
\ref{end-point_z}).
This point would be missed when looking only at the leading twist term of
the \( q\overline{q} \) dipole scattering with two gluon coupling
to the exchange.

The \( \gamma ^{*}\gamma ^{*} \) impact factor shows the same structure
of end point contributions at large s. The amplitude is the one of
non-forward virtual Compton scattering in the Regge asymptotics 
\( x_{1},x_{2}\ll 1 \).
The standard short distance factorization for the Compton amplitude
at \( x_{1},x_{2}=O(1) \) can be continued into this region. In leading
\( \ln Q^{2} \) we have two cases. The hard scattering sub-process
may be the one of Compton scattering off a quark which enters the
amplitude in convolution with the (generalized) quark distribution.
This means, the coefficient functions starts at tree level; the quark
loop contributes only in next-to-leading order. In the other case
the hard scattering sub-process is the one of 
\( \gamma ^{*}g\rightarrow \gamma ^{*}g \)
via a quark loop entering the amplitude in convolution with the (generalized)
gluon distribution; the corresponding coefficient function starts
with the one-loop order. This suggests the way how to treat the end
point contributions in the $\gamma^* V$ case.

We start from the limiting form impact factor at large $Q^2$ from which we
have obtained the end-point contributions (\ref{end-point}) (compare also
Appendix (\ref{photonmeson},\ref{endpointmeson})),
\bea
\label{endfactW}
\Phi^{V \lambda_i \lambda_f}_{z_0} = 
\frac{2}{m_V^2} 
 \int \frac{d^2 \ell^{\prime}}{z} e^{-\frac{|\ell^{\prime}|^2}{z m_{V}^2}}
\int_0^1 dz \cr
\left \{ { W_0^{\lambda_i \lambda_f} (\frac{m_V^2}{Q^2+m_V^2}, \kappa, q) + 
W_1^{\lambda_i \lambda_f} (\frac{m_V^2}{Q^2+m_V^2}, \kappa, q) \frac{1}{z}
\over
Q^2 + q \kappa^* + q^* \kappa + \frac{|\ell^{\prime}|^2}{z} + 
\frac{|\kappa|^2}{z} }
- ...(\kappa + q = 0 ) ...\right \}
\eea  
We have restored the integration over the loop transverse momentum $\ell $
(\ref{photonmesonl}), 
$\ell^{\prime} = \ell - \frac{1}{1+\lambda} (\kappa + z q) 
\approx
\ell - \kappa $. Further we restore the integrations over the loop momentum
Sudakov components $\beta_{\ell} $, (\ref{sudakov}), and over the subenergy
$(q_1 - k_1)^2 \approx \beta_1 s $ by including the mass shell
$\delta$-functions (\ref{mass_shell}).
\be
\label{endfactd}
\Phi^{V \lambda_i \lambda_f}_{z_0} = 
\int_0^1 s d\beta_1 \int_0^{\beta_1} d\beta_{\ell} \ ... \ 
z \delta (z (\beta_1 - \beta_{\ell} ) s - |\ell^{\prime} |^2 )  \ \ \ 
\delta ( (\beta_{\ell} - x_1) s - |\ell^{\prime} + \kappa |^2 )
\ee
The periods stand for the right-hand side of (\ref{endfactW}).
Now the first $\delta$-function is used to do the integral over $z$;
in particular we do the substitution $ \frac{|\ell\p|^2}{z} = (\beta_1 -
\beta_{\ell} ) s $. We notice that the first term in the argument of the
second $\delta$-function dominates for large $Q^2 = x_1 s$. 

We retain only the contributions with a logarithmic contribition of the $z$
integral. If substituted into the amplitude only the structure
$f^{(2)}(\kappa,q)$ results in a logarithmic $\kappa $ integral
allowing to approximate the proton impact factor by the unintegrated
generalized  gluon distribution. In fact we pick up the leading
contribution in the kinematics
$s \gg Q^2 \gg |\ell|^2 \gg |\kappa|^2 \gg |q|^2 \sim m_V^2 $.
\bea
\label{endfactell}
\Phi^{V \lambda_i \lambda_f}_{z_0} = 
\int_0^1 s d\beta_1 \int_0^{\beta_1} d\beta_{\ell 1} 
\int_{|\kappa|^2}^{Q^2} \frac{d^2\ell\p}{|\ell\p|^2 } \ 
\delta(\beta_{\ell 1} - x_1)  \cr
{2 e^{-(\beta_{ 1} - \beta_{\ell 1}) \frac{s}{m_V^2}} \over
\tilde Q^2 + (\beta_{ 1} - \beta_{\ell 1}) s } \ 
\left \{ {W_0 \ |\kappa|^2 \over  \tilde Q^2 + (\beta_{ 1} - \beta_{\ell
1})s }
+ W_1 \right \}.
\eea
The resulting end-point contribution to the amplitude has the form
\bea
\label{endfactGLAP}
M^{V \lambda_i \lambda_f}_{z_0} = C^{V \lambda_i \lambda_f}
\frac{\bar f^{(n)}(q)}{\tilde Q^n} 
\int_{m_V^2}^{Q^2}  \frac{d^2\ell\p}{|\ell\p|^2 }  
\alpha_S (|\ell\p|^2) 
\int_0^1 s d\beta_1 \int_0^{\beta_1} d\beta_{\ell 1} 
\int_{|\kappa|^2}^{Q^2} 
\delta(\beta_{\ell 1} - x_1)  \cr
P^V(\beta_{\ell 1}, \beta_{\ell2};\beta_1, \beta_2  ) \ 
G_g (\beta_1,\beta_2, q; |\ell\p|^2 ), 
\eea
where
$$
\bar f^{(n)} (q) = \int d\varphi_k { f^{(n)} (\kappa, q) \over |\kappa|^2 }.
$$
This is just the evolution term in the integral form of the 
generalized GLAPD equation, corresponding to the parton splitting
$gg \rightarrow q \bar q$. The splitting kernel,
\be
\label{GLAPDkernelV}
P(\beta_{\ell 1},\beta_{\ell 2}; \beta_1,\beta_2) = 
(\beta_{ 1} - \beta_{\ell 1})^{-1} \ \exp (- \frac{\beta_{1} - \beta_{\ell
1})}{\beta_{\ell 2}  } ) \ 
[ 1 + \frac{\beta_{1} - \beta_{\ell 1} }{\beta_{\ell 1} }]^{-1 -a},
\ee
is an unconventional one, 
because the resulting $q \bar q$ state is not the one of
leading twist exchange. 
We have $x_2 = \beta_{\ell 2} = \frac{m_V^2}{s} \ll 1 $
and therefore the kernel results in a narrow distribution peaked around 
$\beta_{\ell 1} = \beta_1$.

\section{Generalized Bjorken evolution}
\setcounter{equation}{0}

We consider in some details the end-point contribution of the \( \gamma ^{*}p\rightarrow \gamma ^{*}p \)
amplitude with \( \lambda _{i}=\lambda _{f}=1. \) We expect to recover
the GLAPD evolution \( gg\rightarrow q\overline{q} \) known from
the case of the structure function \( F_{2} \) and the non-forward
generalization (DVCS).

For simplicity we restrict ourselves to purely longitudinal momentum
transfer, \( q=0,x_{1}\neq x_{2} \). The end-point contribution to
the impact factor reads
\be
\label{end-point_contr_to_IP}
\Phi ^{\gamma ,1,1}(k,0)|_{z_{0}}=
\int d^{2}\ell \int ^{_{z_{0}}}_{0}dz\left\{ \frac{e^2 \alpha_S 
\ell(\ell-\kappa)^{*}
\frac{1}{z^{2}}}{
\left[ x_{1}s+\frac{|\ell|^{2}}{z}\right] \left[ x_{2}s+
\frac{|\ell-\kappa|^{2}}{s}
\right] }-...(\kappa = 0)...\right\} 
\ee
The subtraction term is obtained from the written term by substituting
\( \kappa = 0 \)

\bea
\label{endp11}
\Phi ^{\gamma ,1,1}(k,0)|_{z_{0}}= e^2 \alpha_S
\int \frac{d^{2}\ell^{\prime}}{z}\int ^{1}_{0}dy\int _{0}^{z_{0}}
\frac{dz}{z}\left\{ \frac{|\ell^{\prime }|^{2}-
\overline{y}|\kappa|^{2}}{\left[ s(x_{1}y+x_{2}\overline{y})+
\frac{|\ell^{\prime} |^{2}}{z}+
\frac{|\kappa|^{2}}{z}y\overline{y}\right] ^{2}}
\ -...(\kappa=0)...\right\} 
\cr
=e^2 \alpha_S \int \frac{d^{2}\ell^{\prime} }{z}
\int ^{1}_{0}dy\int _{0}^{z_{0}}
\frac{dz}{z}\left\{ \frac{-2\frac{|\ell^{\prime }|^{2}}
{z}|\kappa|^{2}y\overline{y}
-\overline{y}|\kappa|^{2}s(x_{1}y+x_{2}\overline{y})+
{\cal O}(|\kappa|^{4})}{\left[ s(x_{1}y+x_{2}\overline{y})+
\frac{|\ell^{\prime} |^{2}}{z}\right] ^{3}}\right\} 
\eea
As in (\ref{mass_shell}) we restore the integration 
over \( \beta _{1},\beta _{\ell} \)
by including the mass shell \( \delta  \) function. This allows to
substitute in the integrand
\[
\frac{|\ell^{\prime }|^{2}}{z}=(\beta _{1}-\beta _{l_{1}})s=
(\beta _{2}-\beta _{l_{2}})s,
\ \ \beta _{l_{1}}= x_{1,}\beta _{l_{2}}= x_{2}\]
and in particular\[
s(x_{1}y+x_{2}\overline{y})+\frac{|l^{\prime }|^{2}}{z}=
 s(\beta _{1}y+\beta _{2}y)\]
We obtain in analogy to the previous section 
\bea
\label{step_back}
\Phi ^{\gamma ,1,1}(k,0)|_{z_{0}}= e^2 \frac{|\kappa|^{2}}{s}
\int ^{z_{0}Q^{2}}_{|\kappa|^{2}} \alpha_S (|\ell\p|^2)
\frac{d|\ell^{\prime} |^{2}}{|\ell^{\prime }|^{2}}
\int d\beta _{1}d\beta _{2}d\beta _{l_{1}}
\delta (x_{1}-x_{2}-\beta _{1}+\beta _{2}) \cr
\delta (\beta _{l_{1}}-x_{1})
\int ^{1}_{0}dy\frac{\beta _{1}y+\beta _{2}\overline{y}+
2(\beta _{l_{1}}-\beta _{1})y\overline{y}}
{[\beta _{1}y+\beta _{2}\overline{y}]^{3}}\ \Theta (\beta _{1}-x_{1})
\eea
The result contributes to the amplitude 
\( \gamma ^{*}p\rightarrow \gamma ^{*}p \)
\( (\lambda _{i}=\lambda _{f}=1) \) by convolution with the unintegrated
gluon distribution,
\bea
\label{via_splitting}
M^{\gamma ,11}=e^2 \int ^{z_{0}Q^{2}}_{m_{V}^{2}} \alpha_S (|\ell|^2)
\frac{d|\ell|^{2}}{|\ell|^{2}}
\int ^{1}_{0}d\beta _{1}d\beta _{2}
\delta (x_{1}-x_{2}-\beta _{1}+\beta _{2})
P(x_{1}, x_{2};\beta _{1},\beta _{2}) \cr
G_{g}(\beta _{1},\beta _{2};q=0,|\ell|^{2})
\eea
As expected, the result has the form of the evolution term in the
integral representation of the GLAPD equation with the (non-forward)
splitting kernel
\be
\label{splitf}
P(x_{1,}x_{2};\beta _{1},\beta _{2})=
\frac{\Theta (\beta _{1}-x_{1})}{\beta _{1}\beta _{2}}
(1+ {\cal O} (\beta _{1}-x_{1}))
\ee
We compare \( P(x_{1,}x_{2};\beta _{1},\beta _{2}) \) with the generalized
GLAPD evolution kernel as calculated in \cite{Bukhvostov:rn}
\bea
\label{Bukhvostov}
H(x_{1},x_{2},\beta _{1},\beta _{2})=
\frac{1}{\beta _{1}\beta _{2}}
\left\{ x_{1}(J_{1}-J_{11^{\prime }})+(x_{1}-\beta _{1})J_{1}
+(\beta _{1}\Leftrightarrow \beta _{2})\right\} \cr
J_{1}=\int ^{\infty }_{-\infty }
\frac{d\alpha }{2\pi i}[\alpha x_{1}+1-i\varepsilon ]^{-1}
[- \alpha x_{2}+1-i\varepsilon ]^{-1}
[\alpha (x_{1}-\beta _{1})+1-i\varepsilon ]^{-1} \cr
J_{11^{\prime }}=\int ^{\infty }_{-\infty }
\frac{d\alpha }{2\pi i}[\alpha x_{1}+1-i\varepsilon ]^{-1}
[\alpha (x_{1}-\beta _{1})+1-i\varepsilon ]^{-1}
\eea
The first (second) term in the bracket corresponds to the \( gg\rightarrow qq \)
transition without(with) spin flip.
For the impact factor the right-hand cut discontinuity in \( \beta _{1}s \)
is relevant
\bea
\label{discontinuity}
disc_{\beta _{1}}H=\frac{1}{\beta _{1}\beta _{2}}
\left\{ \frac{x_{1}}{\beta _{1}}\frac{x_{2}}{\beta _{2}}
+\frac{(x_{1}-\beta _{1})(x_{2}-\beta _{2})}{\beta _{1}\beta _{2}}\right\} 
\Theta (\beta _{1}-x_{1}) \cr
=P(x_{1},x_{2};\beta _{1}\beta _{2})
\eea
The familiar forward splitting kernel for \( gg\rightarrow qq \)
is recovered for \( \beta _{1}=\beta _{2},x_{1}=x_{2} \).

The comparison of (5.7) and (5.5) confirms that the end-point contribution
to the diffractive amplitude written in impact factor representation
is to
be absorbed into the evolution of the generalized gluon distribution.
Since the impact factor ansatz accounts for the Regge asymptotics
the corresponding evolution contribution results merely in this approximation;
in particular the subenergy \( (Q_{1}-k_{1})^{2}\simeq s(\beta _{1}-x_{1}) \)
is small in this asymptotics.

\section{Amplitudes and numerical estimates}

The decomposition of the $z$ range in the impact factor
(\ref{division}) results in the amplitude as a sum of corresponding 3
terms, $ M^{\lambda_i, \lambda_f} =
  M_1^{\lambda_i, \lambda_f} +
  M_{z_0}^{\lambda_i, \lambda_f} +
  M_{\bar z_0}^{\lambda_i, \lambda_f} $.
 According to (\ref{leading_asymp}) the contribution from $z = {\cal O}
(1)$ is calculated as
\be
 M_1^{\lambda_i, \lambda_f} = \int_{m_V^2}^{Q^2} d^2 \kappa \
\int_{z_0}^{\bar z_0} z \bar z \varphi_4^{\lambda_i \lambda_f} (q,
\kappa) dz {1 \over |\kappa|^2 |q+\kappa|^2 }
G_g^{\prime} (x_1,x_2,q, |\kappa|^2).
\ee

The leading $\ln Q^2$ contribution of the integral over $\kappa $
results in
\be
  M_1^{\lambda_i, \lambda_f} \approx \left (
\int_{z_0}^{\bar z_0} z \bar z \bar \varphi_4^{\lambda_i \lambda_f}
dz \right )
\cdot G_g (x_1,x_2,q;Q^2).
\ee

Here $ \bar \varphi_4^{\lambda_i \lambda_f} $ are the coefficients in
(\ref{main}) accompanying $f^{(2)} (\kappa, q)$ as in
(\ref{leading_impact}). For example, in the case $\lambda_i = \lambda_f
= 1$ we have
\be
  M_1^{1 1} \approx \left (
\int_{z_0}^{\bar z_0} (\frac{1}{z \bar z} - 2 ) \
dz \right ) {C^{V 11} \over Q^2 +m_V^2 }
\cdot G_g (x_1,x_2,q;Q^2).
\ee
The end point contributions are not small with $z_0$ for the terms in
(\ref{main}) involving $\frac{1}{z \bar z}$. As explained in Section 4
the logarithmic $z$ integral is a generalized Bjorken evolution
(\ref{endfactGLAP}) term resulting in the effective quark distribution
\be
\label{effquark}
\tilde G_q (x_1,x_2, q, z_0 Q^2) =
\int_{m_V^2}^{z_0 Q^2} {d|\ell |^2 \over |\ell|^2}
{\alpha_S(|\ell |^2 ) \over \alpha_S(Q^2) }
G_g (x_1,x_2, q, |\ell|^2).
\ee
In the example $\lambda_i = \lambda_f = 1 $ this leads to
\be
M_{z_0}^{ 1 1 } = M_{\bar z_0}^{1 1} =
 {C^{V 11} \over Q^2 +m_V^2 } \cdot
\tilde G_q (x_1,x_2, q, z_0 Q^2)
\ee

The sum of the three contributions to the amplitude has the form
(\ref{leading_ampl}).
Notice that the effective quark distribution has been indentified with
the Bjorken evolution term of the gluon distribution. This involves the
reasonable assumption that this distribution is small for small $x$ 
at the scale $m_V^2$.

The generalized gluon distribution at small $x_{1/2}$  is
adopted here to be proportional to the ordinary small $x$ gluon distribution
\be
 G_g (x_1,x_2, q,  Q^2) = c  \ G_g (\frac{x_1+x_2}{2}, Q^2 ) \ e^{-b |q|^2}
\ee
For the slope parameter we adopt the value $b = 6 \ GeV^{-2} $.

The generalized Bjorken evolution leads to a modification of this relation
if it is assumed to hold at some $Q_0^2$. This effect has been investigated
for small $x$ and small skewedness \cite{Shuvaev:1999ce} and has been 
 applied to diffractive vector meson production \cite{Martin:1999wb}.
In the latter paper the $Q^2$ dependence of the cross sections 
is found to be changed by
50 per cent for light vector mesons  at the highest $Q^2 \  ( = 30 GeV^2)$. 
Since the aim of the numerical estimates in this paper  
is merely the illustration of
the proposed factorization concept, we do not include
the skewdness effects here for simplicity, understanding that this 
would be an appropriate further improvement.

Further, for the numerical estimate we have to specify the gluon
distribution. Although our approach is based on the asymptotic expansion
in $Q^2$ for comparison with the data we have to extrapolate the results
to non-asymptotic values of this scale. Moreover, in the reconstruction of
the effective quark distribution according to (\ref{effquark}) we have
to integrate the small $x$ gluon distribution 
starting from $m_V^2 \approx .5 GeV^2 $.
This means we need the gluon distribution at small scales, where
the standard parametrizations like MRST do not give a certain answer
and where the application of the evolution equation is not reliable.
Therefore we adopt the two-pomeron paramerization \cite{Donnachie:2001xx},

\[ {G_g}(x,Q^{2})\sim X_{0}\left( \frac{Q^{2}}{1+
\frac{Q^{2}}{Q^{2}_{0}}}\right) ^{1+\epsilon _{0}}(1+
\frac{Q^{2}}{Q^{2}_{0}})^{\varepsilon _{0}/2}
x^{-\varepsilon _{0}-1}+X_{1}(1+\frac{Q^{2}}
{Q^{2}_{1}})^{\varepsilon _{1}/2}x^{-\varepsilon _{1}-1},\]

where \( \varepsilon _{0}=0.43 \), \( \epsilon _{1}=0.08 \), \( X_{0}=0.0014 \),
\( X_{1}=0.5954, \) \( Q^{2}_{0}=9.108 \), \( Q^{2}_{1}=0.5894 \).

Again, we have no particular reason to favour this parametrization
besides of its convenience in our aim of illustration.
We did not try to optimize the results with regard to different
possible input parametrizations.

The cut-off $z_0$ is to separate the $z$ range of order unity from the
end-point regions; a value $z_0 = 0.1 ... 0.2$ is reasonable and indeed
in this range the $z_0$ dependence of the result is weak.
The condition that $z_0 Q^2 $ is much larger than $m_V^2$ leads to the
restriction of the applicability of the estimates to $Q^2 > 5 GeV^2 $.

We have calculated in the given approximation with the input specified above
the $Q^2$ dependence of the ratio of longitudinal to transverse virtual photon
diffractive cross sections. We do this  including all helicity
amplitudes $(R)$ and also with the helicity conserving amplitudes only
($R_0$), the latter case corresponds to the ratio usually obtained in the
data analysis.
Thus we calculate
\be
R(Q^2) = {\sigma_L (Q^2) \over \sigma_T (Q^2) }, \ \ 
R_0(Q^2) = {\sigma_L^{(0)} (Q^2) \over \sigma_T^{(0)} (Q^2) },
\ee
where
\bea
\sigma_L (Q^2) = \int dt (|M^{00}|^2 + 2 |M^{01}|^2), \ \ 
\sigma_L^{(0)} (Q^2) = \int dt |M^{00}|^2, \cr
\sigma_T (Q^2) = \int dt (|M^{11}|^2 +  |M^{10}|^2 + |M^{1 -1}|^2 ), \ \ 
\sigma_T^{(0)} (Q^2) = \int dt |M^{11}|^2.
\eea
The results are shown in Fig. 2 together with HERA data. $R_0$ deviates from
$R$ by 10 per cent at higher $Q^2$. The inclusion of the next-to-leading terms
in the twist expansion does not lead to  noticeable changes.

\begin{figure}[htbp]
\begin{center}
\vspace{-0.cm}

\epsfig{file=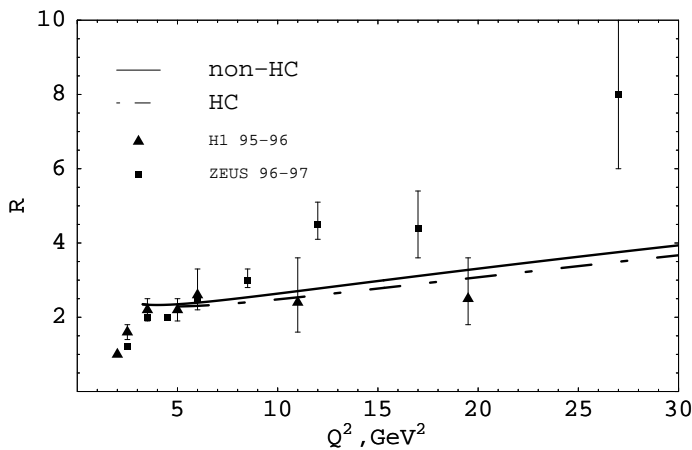,width=10cm}
\end{center}
\caption{Ratio of the longitudinal and transverse 
elastic $\rho^{0}$ electroproduction cross sections as a function of 
$Q^{2}$. The dotted line corresponds to the assumption of helicity conservation,
the solid line takes into account spin flip amplitudes.}
\label{fig:R}
\end{figure}

The coefficients in the angular-decay distribution are more sensitive to the
helicity dependence, because in some of them the small flip amplitudes enter
in the first power. We use the following expressions for the coefficients 
$r^{\alpha}_{ik}$ in terms of the helicity amplitudes $M^{\lambda_i
\lambda_f}$
simplified in comparison to \cite{Schilling:1973ag}
for the appropriate case of the virtual photon polarization
parameters being $\epsilon \approx 1, \delta \approx 0$. 

\[
r^{04}_{00}\propto \frac{1}{N}(|M^{00}|^{2}+|M^{10}|^{2})\]
\[
r_{00}^{5}\propto \frac{1}{N}Re(M^{00 *} M^{10})\]
\[
r^{5}_{11}\propto (Re(M^{01}M ^{* 11})-Re(M^{01}M ^{* 1-1}))\]
\[
r^{1}_{00}\propto -\frac{1}{N}|M^{10}|^{2}\]
\[
r^{1}_{11}\propto \frac{1}{N}(M^{1-1}M ^{* 11}+M^{11}M^{* 1-1}),\]
\[
N=|M ^{00}|^{2}+|M ^{10}|^{2}+ 2 |M^{01}|^{2}+|M^{11}|^{2}+|M^{1-1}|^{2}\]

Results are given for \( r^{04}_{00} \), \( r^{1}_{00}+2r^{5}_{11}, \)
\( r^{1}_{00}+2r^{1}_{11} \). In the data analysis the first coefficient is
extracted from the
dependence on the polar angle of the $\pi^+$ in the vector meson
rest frame with respect to the vector meson momentum direction.
Actually $R_0$ is obtained from this coefficient by a formula valid for the 
ratio in the approximation of helicity conservation. 
The other combinations are the ones obtained in the data analysis from the
dependence on the angle between the lepton scattering and vector meson
production planes.

The  $t$ dependence of the coefficients $r^{\alpha}_{ij}$ has been
calculated at a fixed value of $Q^2 = 10 GeV^2$. The results are shown in 
Fig. 3 in comparison with the data for a broad  range $Q^2 = 3  ... 30 \ GeV^2 $
 with the average
values of $6.5 \ GeV^2$ for the ZEUS data points and $5 \ GeV^2$ 
for the H1 data points.

\begin{figure}[htbp]
\begin{center}
\vspace{-0.cm}

\epsfig{file=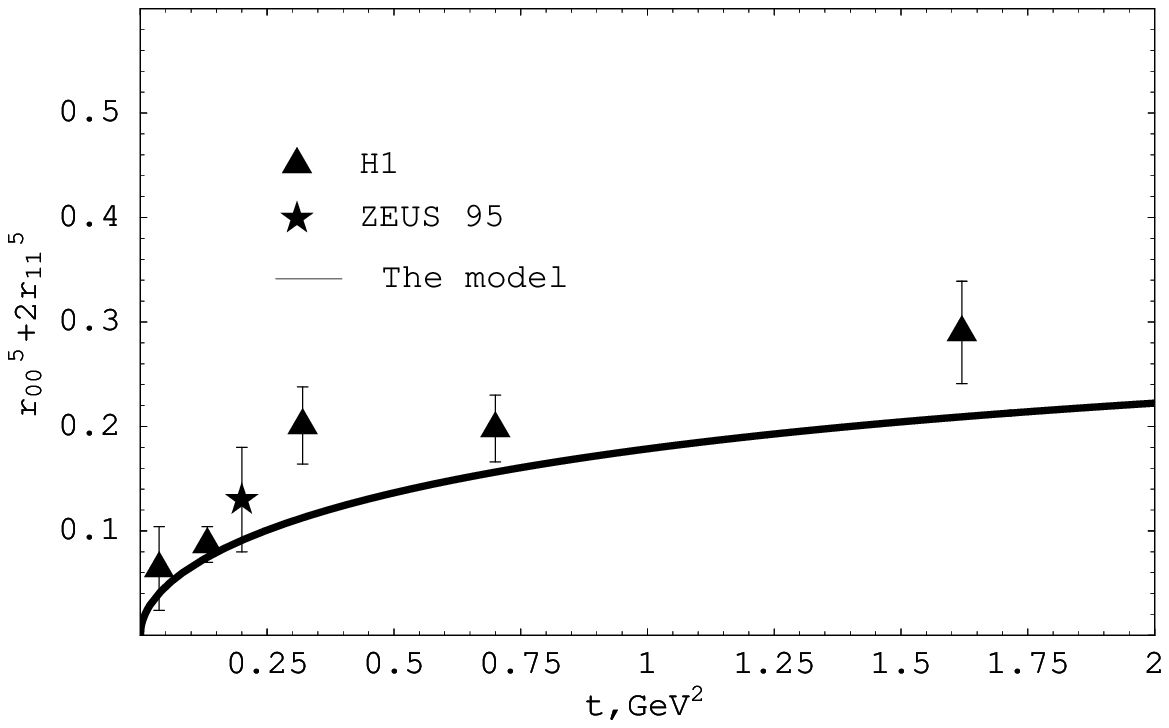,width=7.9cm}
\epsfig{file=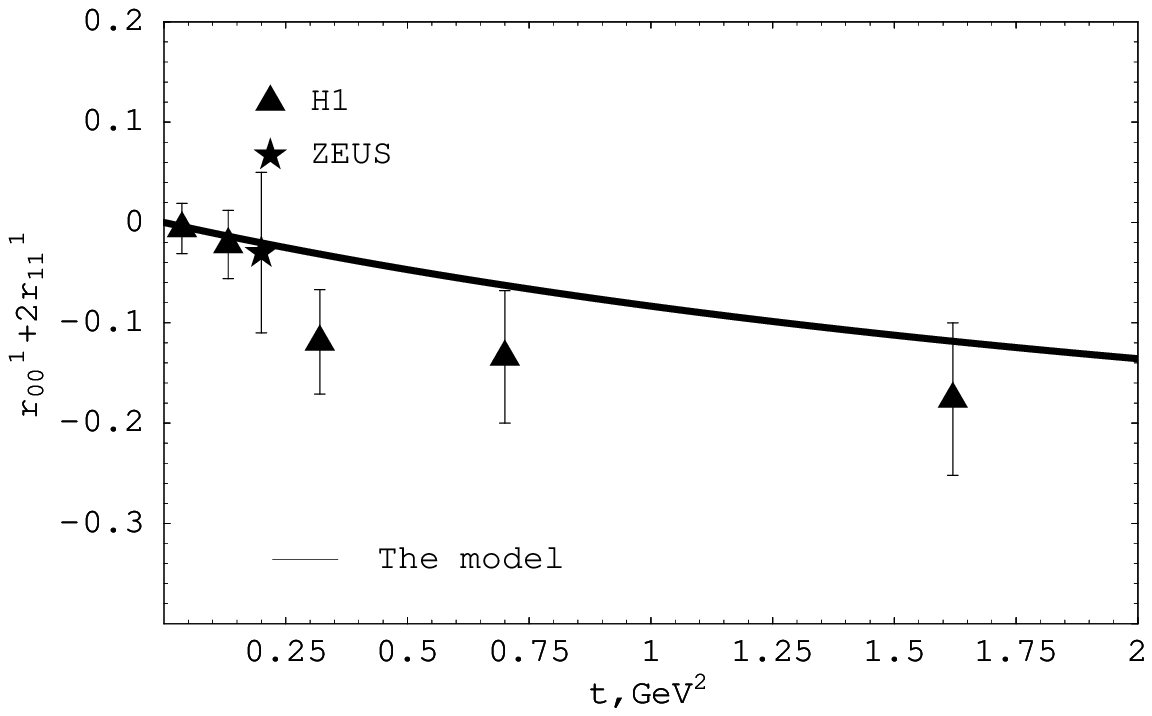,width=7.9cm}
\epsfig{file=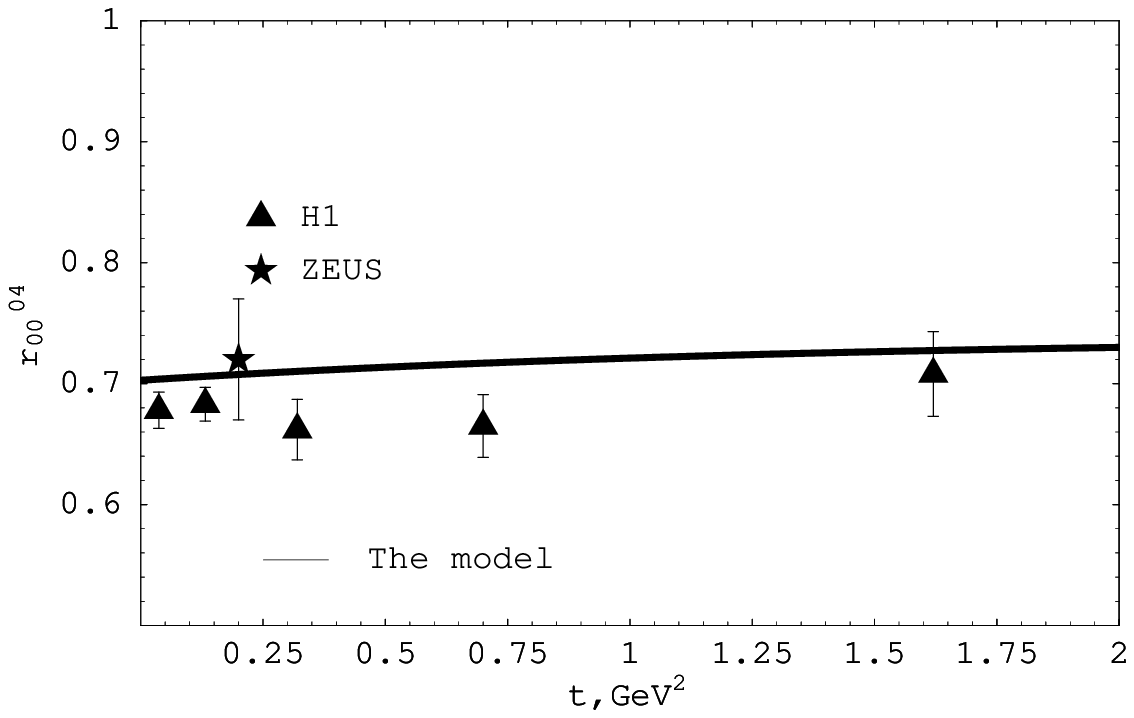,width=7.5cm}
\end{center}
\caption{$r_{00}^5+2r_{11}^5$, $r_{00}^1+2r_{11}^1$, $r_{00}^{04}$ 
as a function of transferred momentum t. }
\label{fig:rt}
\end{figure}

In evaluting the $Q^2$
dependence we have substituted in the amplitudes the average value
of $t$ determined by the slope parameter $b$.
The results are shown in Fig. 4.

 \begin{figure}[htbp]
\begin{center}
\vspace{-0.cm}

\epsfig{file=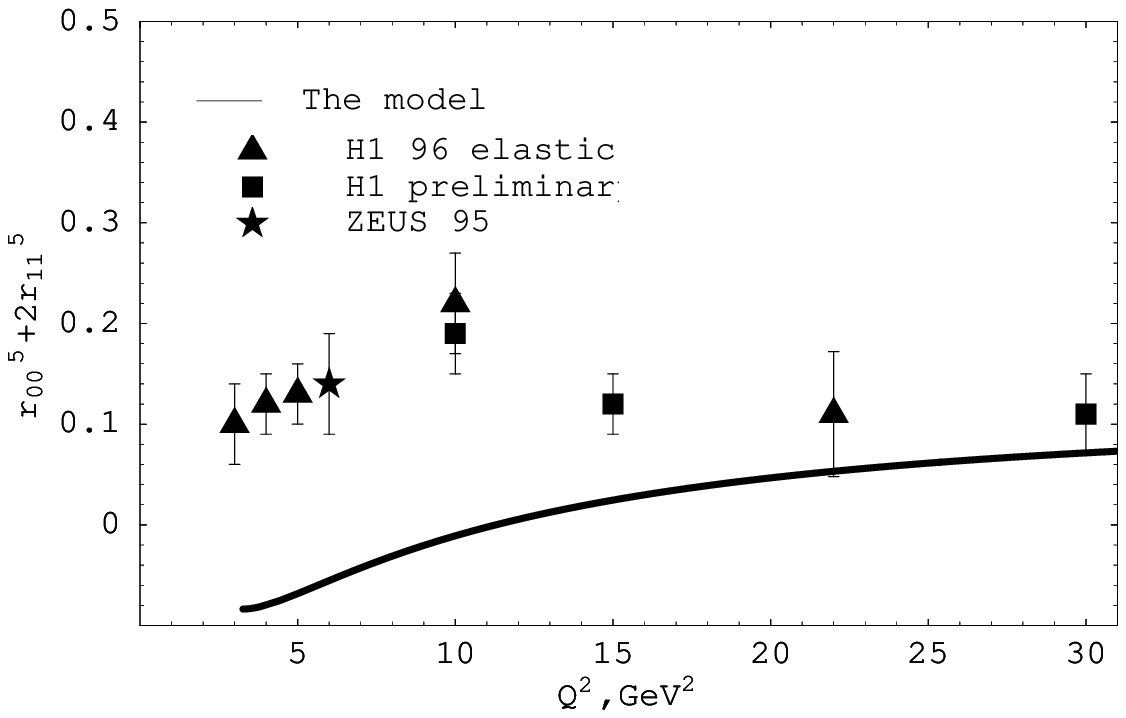,width=7.9cm}
\epsfig{file=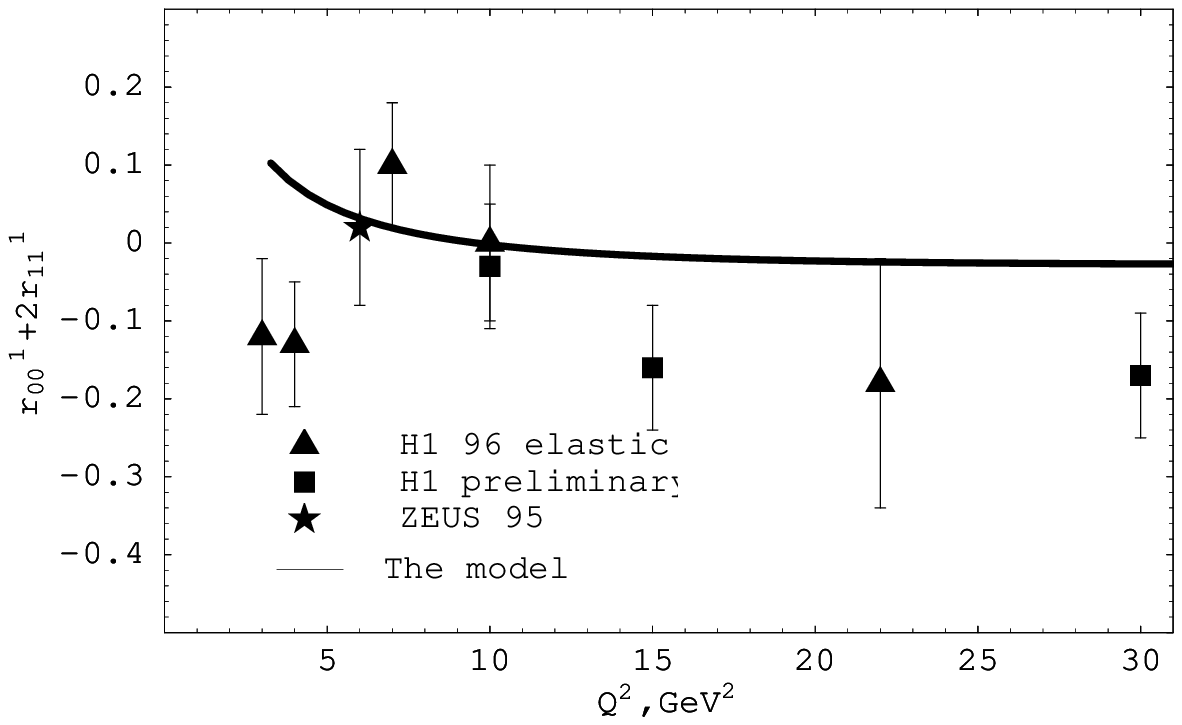,width=7.9cm}
\epsfig{file=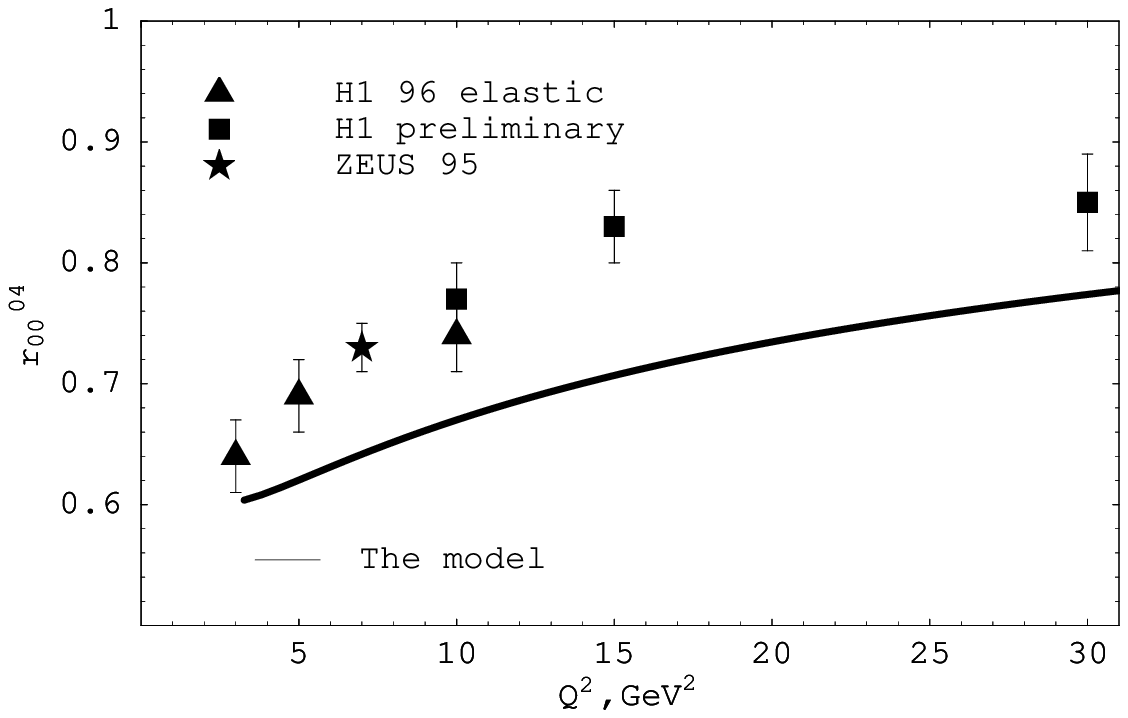,width=7.5cm}
\end{center}
\caption{$r_{00}^5+2r_{11}^5$, $r_{00}^1+2r_{11}^1$, $r_{00}^{04}$
as a function of $Q^2$.}
\label{fig:rq}
\end{figure}

\section{Discussion}

The diffractive $\gamma^* V$ amplitudes constructed with a $q \bar q$ dipole
impact factor involve enhanced contributions from the end-point regions
in the quark longitudinal momentum fraction $z$, besides of the leading
contribution to the $\lambda_i = \lambda_f =0 $ helicity amplitude. These
endpoint contributions appear as singularities in the extrapolation of the
twist expansion . However, resummed corrections proportional to 
$(\frac{m_V^2}{z \bar z Q^2})^n $  
remove the singularities at $z, \bar z = 0$ and result in
logarithmically enhanced end-point contributions. The latter are shown to
be the ones of the generalized Bjorken evolution of the $t$-channel parton
exchange $gg \rightarrow q \bar q$. The short-distance factorization of the
diffractive amplitude in terms of the $\gamma^* V$ transition and the
Pomeron exchange therefore involves contributions of two types: one with a
scattering $q \bar q$ dipole coupled by $gg$ to the Pomeron and one with a
(Compton like) scattering  $q$ or $ \bar q$ coupled by $q \bar q$ to the
Pomeron. The $q \bar q$ exchange differs from the leading twist one. 

We
have chosen the impact factor form of the diffractive ampltude as a starting
point. Our lack of understanding of the proton impact factor and of the
Pomeron coupling to it has been managed by replacing their convolution by
the unintegrated generalized parton distribution, a standard step in 
$k_T$ factorization schemes. This is justified in so far as it results at
large $Q^2$ in the parton distribution convoluted with the coefficient
function resulting from the $\gamma^* V$ impact factor. Notice that in the
leading $\ln Q^2$ approximation contributions of the impact factors
depending on the azimuthal angle of the exchanged parton momentum, 
$\frac{\kappa}{|\kappa |}$, drop out. We see here a possible source of
corrections, which may be relevant at moderate $Q^2$. 

The resulting large $Q^2$ asymptotics of the amplitudes  consists of terms
with the generalized small $x$ gluon distribution and with the effective $q
\bar q $ distribution. The latter appears as an additional non-perturbative
input. However, its main contribution results from the evolution of the
gluon distribution with a splitting function transferring the longitudinal
momentum fraction from $g$ to $q$ almost unchanged. 

In this way we see that the construction of the diffractive amplitude with a
scale of the gluon distribution replaced by $ z \bar z Q^2$ 
\cite{Martin:1996bp}
is actually an approximation to the factorization proposed here, because
with that scale replacement the end-point contributions of the
$z$-integration are approximately the $q \bar q$ exchange contribution
by $gg \rightarrow q \bar q$ evolution. This scale replacement has been
applied in a previous study of the polarization effects \cite{Ivanov:1998gk}.

In our construction the terms $\sim (\frac{m_V^2}{z \bar z Q^2} )^n$
improving the end-point behaviour are introduced via the vector meson light
cone wave function. It is modelled starting from the perturbative $\gamma^*$ 
wave function by keeping its helicity structure and removing its hard
contributions. Wave functions sharing these basic features lead to similar
results for the large $Q^2$ helicity amplitudes. In the present scheme the
higher twist correction to the amplitudes can be calculated. The leading
asymptotics involves the value of the impact parameter wave function at the
origin and this can be recast in  of the distribution amplitude
formulation. 

We did not include the hard scale evolution of the wave
function \cite{ERBL}. In the numerical estimates we did not include the
skewedness effects. We have used a particular parametrization of the 
small $x$ gluon distribution covering the small $Q^2$ range needed in
particular for reconstructing the effective $q \bar q$ distribution
therefrom.

The results for the cross  section ratio and the Schilling-Wolf coefficients
have been calculated using  this input with no attempts of fitting.
We have shown that they describe the basic features of the data.

 Including the mentioned improvements the proposed factorization scheme 
provides a theoretical framework applicable to diffractive electroproduction
free of  previous uncertainties about the end-point contributions and
therefore useful for extracting information about the structure of the
hadrons involved from data analysis.
We expect that the scheme applies to other hard diffractive processes.

\vspace{2cm}

{\Large \bf Acknowledgements }

The authors  thank D.Yu. Ivanov, B. Pire and L. Szymanowski 
for discussions. 
One of us (A.I.) is fellow of the Leipzig
Graduate College "Quantum Field Theory" supported by Deutsche
Forschungsgemeinschaft.

\section{Appendix}
\setcounter{equation}{0}

We take the simple form of (\ref{dipole}) into account
and write
\begin{equation}
\label{IF}
\Phi (k,q)=\int ^{1}_{0}dzz\overline{z}[\varphi (k,z,q)+
\varphi (-k-q,z,q)-
\varphi (0,z,q)-\varphi (-q,z,q) ],
\end{equation}

\[
\varphi (k,z,q)=\int d^{2}\ell_{1}d^{2}\ell_{2}\delta (\ell_{2}-\ell_{1}-
\kappa - zq)
\Psi ^{\lambda _{i}}_{i}(\ell_{1})\Psi ^{*\lambda _{f}}_{f}(\ell_{2})\]

For the
\( \gamma ^{*}\gamma ^{*} \) case we have
\be
\label{photonphotonl}
\varphi^{\gamma}  (\kappa, q,z) = 
\int d^2 \ell { e^2 \alpha_S V^{\lambda_i} V^{ \lambda_f *}  \over
[Q_1^2 + \frac{|\ell|^2 + m_q^2}{z \bar z} ]
[Q_2^2 + \frac{|\ell -\kappa -z q|^2 + m_q^2}{z \bar z} ] }
\ee
Doing the integration over the transverse momentum we obtain 
\be
\label{photonphoton}
\varphi ^{\gamma }(\kappa,z,q)= e^2 \alpha_S 
\pi \int ^{1}_{0}\frac{dy}{y\overline{y}}
\frac{\left\langle V^{\lambda _{i}}_{i}V^{\lambda _{f} *}_{f}
\right\rangle ^{\prime }(y,Q_{1},Q_{2})}{\left[ \frac{Q^{2}_{1}}{\overline{y}}
+
\frac{Q^{2}_{2}}{y}+\frac{m^{2}_{q}}{z\overline{z}y\overline{y}}+
\frac{|\kappa + zq|^{2}}{z\overline{z}}\right] }
\ee

For the \( \gamma ^{*}V \) case we have
\bea
\label{photonmesonl}
\varphi^{V}  (\kappa, q,z) = 
\int d^2 \ell_2 
\exp(- \frac{|\ell_2|^2 +m_q^2}{m_V^2 z \bar z } ) \cr
{ e f_V \alpha_S V^{\lambda_i} V^{ \lambda_f *} 
 \over
[Q^2 + \frac{|\ell_2 - \kappa -z q|^2 + m_q^2}{z \bar z} ] m_V^2 }.
\eea
The result of the integration over $\ell_2$ can be written as
\be
\label{photonmeson}
\varphi ^{V}(\kappa ,z,q)=e f_V \alpha_S 
\pi \int ^{1}_{0}\frac{d\lambda }{1+\lambda }
\frac{\left\langle V^{\lambda _{i}}_{i}
V^{\lambda _{f} *}_{f}\right\rangle ^{\prime }
(\frac{\lambda }{1+\lambda },Q_{1},m_{V})}{m^{2}_{V}}
\ee
\[
\exp[-\frac{Q^{2}}{m^{2}_{V}}\lambda -
\frac{m^{2}_{q}(1+\lambda )}{z\overline{z}m^{2}_{V}}-
\frac{|\kappa+zq|^{2}}{m^{2}_{V}z\overline{z}}\frac{\lambda }{1+\lambda }]\]

We list the numerator \( \left\langle V^{\lambda _{i}}_{i}V^{*\lambda _{f}}_{f}\right\rangle ^{\prime }(y,Q_{1},Q_{2}) \)
for the representative cases of helicities:

\begin{center}
\begin{tabular}{c|c|cc}
\hline 
\( \lambda _{i} \)&
\( \lambda _{f} \)&
\ \ \ \ \ &
\( \left\langle V^{\lambda _{i}}_{i}V^{*\lambda _{f}}_{f}\right
\rangle ^{\prime }(y,Q_{1},Q_{2}) \)\\
\hline
0&
0& & 
\( 2Q_{1}Q_{2} \)\\
0&
-1& &
\( -Q_{1}y(\kappa+zq)(\frac{1}{z}-\frac{1}{\overline{z}}) \)\\
1&
0& &
\( Q_{2}\overline{y}(\kappa+zq)^{*}(\frac{1}{z}-\frac{1}{\overline{z}}) \)\\
1&
-1& &
\( y\overline{y}(\kappa+zq)^{*}\frac{2}{z\overline{z}} \)\\
1&
1& &
\( -\left[ |\kappa+zq|^{2}\overline{y}+z\overline{z}Q^{2}_{1}\right] 
(\frac{1}{z^{2}}+\frac{1}{\overline{z}^{2}}) \)\\
\end{tabular}
\end{center}

As expected, the resulting expressions obey 
\be
\label{symm}
\varphi (\kappa,z,q)=\varphi (-\kappa-q,\overline{z},q)
\ee
and in the integral (\ref{IF}) the number of terms can be reduced
to two.

We calculate the asymptotics of the integral at first for \( z={\cal O}(1) \)
for the \( \gamma ^{*}\gamma ^{*} \)case in the region 
\( s\rightarrow \infty ,Q^{2}_{i}=sx_{i} \), with fixed  $ x_{i}\ll 1 $
\bea
\label{as1}
\varphi ^{\gamma }(k,z,q)=e^2 \alpha_S \pi 
\int ^{1}_{0}dy\left\langle V^{\lambda _{i}}_{i}V^{*\lambda _{f}}_{f}
\right\rangle ^{\prime }(y,sx_{1},sx_{2}) \cr
\left\{ \frac{1}{s}(x_{1}y+x_{2}\overline{y})^{-1}-
\frac{1}{s^{2}}\frac{|\kappa+zq|^{2}y\overline{y}}
{(x_{1}y+x_{2}\overline{y})^{2}z\overline{z}}+
{\cal O} (\frac{1}{s^{3}})\right\} 
\eea
and for the case \( \gamma ^{*}V \) in the region 
\( Q^{2}\rightarrow \infty  \)
\bea
\label{as2}
\varphi ^{V}(\kappa,z,q)\approx e f_V \alpha_S \pi 
\left\langle V^{\lambda _{i}}_{i}V^{*\lambda _{f}}_{f}
\right\rangle ^{\prime }(\frac{m^{2}_{V}}{Q^{2}+m^{2}_{V}},Q_{1},m_{V})
\cr
\ \left\{ \frac{1}{Q^{2}+m_V^{2}}-
\frac{|\kappa+zq|^{2}}{(Q^{2}+m^{2}_{V})z\overline{z}}
+{\cal O} (\frac{1}{(Q^{2}+m_V^{2})^{3}})\right\} 
\eea
Notice that the numerator includes terms proportional to \( s \)
or \( Q^{2} \) if \( \lambda _{i}=\lambda _{f} \). A further term
in the expansion has to be included, leading to cancellation if \( \lambda _{i}=\lambda _{f}=\pm 1. \)
In the case \( \gamma ^{*}V \) it is appropriate to choose the expansion
parameter as \( \frac{m^{2}_{v}}{Q^{2}+m^{2}_{v}} \) resulting in
an improvement of the extrapolation in the region of moderate 
\( Q^{2}>  m^{2}_{V} \).

We obtain in the case \( \gamma ^{*}\gamma ^{*} \)
\be
\label{main_photon}
\varphi ^{\gamma ,00}(\kappa,z,q)z\overline{z}=
C_{1}^{\gamma ,00}(x_{1},x_{2})\frac{|\kappa+zq|^{2}}{s}+
C_{2}^{\gamma ,00}(x_{1},x_{2})\frac{|\kappa+zq|^{4}}{s^{2}}\frac{1}
{z\overline{z}}
+{\cal O} (\frac{1}{s^{3}})
\ee
\[
\varphi ^{\gamma ,01}(\kappa,z,q)z\overline{z}=
C_{1}^{\gamma ,01}(x_{1},x_{2})\frac{|\kappa+zq|^{2}(k+zq)}
{s^{3/2}}(\frac{1}{z}-\frac{1}{\overline{z}})+O(\frac{1}{s^{3}})\]
\[
\varphi ^{\gamma ,10}(\kappa,z,q)z\overline{z}=
C_{1}^{\gamma ,10}(x_{1},x_{2})\frac{|\kappa+zq|^{2}(\kappa+zq)^{*}}{s^{3/2}}
(\frac{1}{z}-\frac{1}{\overline{z}})+{\cal O}(\frac{1}{s^{3}})\]
\[
\varphi ^{\gamma ,1-1}(\kappa,z,q)z\overline{z}=
C_{1}^{\gamma ,1-1}(x_{1},x_{2})\frac{(\kappa+zq)^{*2}}{s}+\]
\[
C_{2}^{\gamma ,1-1}(x_{1},x_{2})
\frac{(k+zq)^{*2}|\kappa+zq|^{2}}{s^{2}}\frac{1}{z\overline{z}}+
{\cal O} (\frac{1}{s^{3}})\]
\[
\varphi ^{\gamma ,11}(\kappa,z,q)z\overline{z}=
C_{1}^{\gamma ,11}(x_{1},x_{2})\frac{|\kappa+zq|^{2}}
{s}z\overline{z}(\frac{1}{z^{2}}+\frac{1}{\overline{z}^{2}})+
{\cal O} (\frac{1}{s^{3}})\]

We use the abbreviation for $C^{\gamma}$ (\ref{photonC}). In the case
$\gamma^* V$ we have similar fomulae. 

When calculating the integrand of (\ref{IF}), 
\be
\varphi _{4}=\varphi (\kappa,z,q)+\varphi (-\kappa-q,z,q)-
\varphi (0,z,q)-\varphi (-q,z,q),
\ee
the dependence on $z$ and $\kappa,\kappa+q$, disentangles. 
We list the relevant expressions
in the following table

\begin{center}
\begin{tabular}{|c|c|}
\hline 
\( \varphi (\kappa,z,q) \)&
\( \varphi _{4}(\kappa,z,q) \)\\
\hline
\multicolumn{1}{|c|}{\( |\kappa+zq|^{2} \) }&
\( \kappa(\kappa+q)^{*}+\kappa^{*}(\kappa+q)=f^{(2)}(k,q) \)\\
\multicolumn{1}{|c|}{\( (\kappa+zq)|\kappa+zq|^{2} \)}&
\( (z - \overline{z}) (q\kappa(\kappa+q)^{*}+q\kappa^{*}(\kappa+q))=
(z-\overline{z})f^{(3)}(\kappa,q) \)\\
\multicolumn{1}{|c|}{\( (\kappa+zq)^{*2} \)}&
\( 2\kappa^{*}(\kappa+q)^{*}=f^{(2,**)}(\kappa,q) \)\\
\multicolumn{1}{|c|}{\( (\kappa+zq)^{* 2}|\kappa+zq|^{2} \)}&
\( \frac{1}{2}f^{(2)}f^{(2,**)}+
(1-3z\overline{z})[q^{*2}f^{(2)}+|q|^{2}f^{(2,**)}] \)\\
\multicolumn{1}{|c|}{\( |\kappa+zq|^{4} \)}&
\( \frac{1}{4}|f^{(2,**)}|^{2}+(1-4z\overline{z})
[|q|^{2}f^{(2)}+\frac{1}{2}q^{*2}f^{(2,**)*}+\frac{1}{2}q^{2}f^{(2,**)}] \)\\
\end{tabular}
\end{center}

\vspace{.5cm}

In this way we obtain the twist expansion valid for \( z=O(1) \).
The result for the case \( \gamma^* V \) is given in eq.(\ref{main}); the
one for \( \gamma ^{*}\gamma ^{*} \) can be recovered by substituting
\( \frac{f^{(n)}}{(Q^{2}+m^{2}_{V})^{n/2}} \) by \( \frac{f^{(n)}}{s^{n/2}} \)
and \( C^{V,\lambda _{i},\lambda _{f}} \) by \( C^{\gamma ,\lambda _{i},\lambda _{f}} \). 

We calculate now the end point contribution to the asymptotic expansion
in \( s \) for \( \gamma ^{*}\gamma ^{*} \) case,
\bea
\label{endpointphoton}
\int ^{z_{0}}_{0}z\overline{z}\varphi ^{\gamma }(\kappa,z,q)dz=
 \int ^{1}_{0}dy\int ^{z_0}_{0}dz \cr
\frac{W_{0}(y,\kappa)+W_{1}(y,\kappa)\frac{1}{z}}{[s(x_{1}y+x_{2}\overline{y})
+y\overline{y}(\kappa q^{*}+\kappa^{*}q+|\kappa|^{2})+
\frac{|\kappa|^{2}y\overline{y}}{z}]} 
+{\cal O} (z_{0})
\eea
and in \( Q^{2} \) for the \( \gamma ^{*}V \) case
\bea
\label{endpointmeson}
\int ^{z_{0}}_{0}z\overline{z}\varphi ^{V}(\kappa,z,q)dz=
 \int ^{z0}_{0}dz\frac{W_{0}(\frac{m^{2}_{V}}
{Q^{2}+m^{2}},\kappa)+W_{1}(\frac{m^{2}_{V}}{Q^{2}+m^{2}},\kappa)
\frac{1}{z}}{[Q^{2}+m^{2}_{V}+(\kappa q^{*}+\kappa^{*} q+|\kappa|^{2})+
\frac{|\kappa|^{2}}{z}]}
\cr
+{\cal O} (z_{0})
\eea
The numerator results from the expansion of 
\( z\overline{z}<V_{i}V_{f}>^{\prime } \)
for small \( z \). \( W_{1} \) is non-vanishing only in the case
\( \lambda _{i}=\lambda _{f}=\pm 1. \) In the case \(  \)\( \lambda _{i}=\lambda _{f}=0 \)
both \( W_{0} \) and \( W_{1} \) vanish, here the small $z$ expansion in the
numerator  starts with \( W_{-1}\sim z \).

Thus we are lead to calculate the integrals 
\be
\label{ep_integral}
I_{a}(\kappa,Q^{2})=\int ^{z_{0}}_{0}dz\frac{z^{-a}}{[Q^{2}+
\frac{|\kappa|^{2}}{z}]}
\ee
\[
=(-\frac{|\kappa|^{2}}{Q^{2}})^{-a+1}Q^{-2}
\ln\frac{Q^{2}z_{0}+|\kappa|^{2}}{|\kappa|^{2}}+{\cal O}(z_{0})\]
and obtain\[
\int ^{z_{0}}_{0}z\overline{z}\varphi ^{V}(\kappa,z,q)dz=
W_{1}(y,\kappa)I_{1}(\kappa,\widetilde{Q}^{2})+
W_{0}(y,\kappa)I_{0}(\kappa,\widetilde{Q}^{2})+
{\cal O} (z_{0})\]
In the case \( \gamma ^{*}\gamma ^{*} \) we have to substitute 
\( \widetilde{Q}^{2} \)
by \( s(x_{1}y+x_{2}\overline{y})+y\overline{y}(\kappa q^{*}+\kappa^{*}q) \);
in the vector-meson case we substitute \( y \) 
by \( \frac{m^{2}_{V}}{Q^{2}+m^{2}_{V}} \)
and \( \widetilde{Q}^{2} \) by \( Q^{2}+m^{2}_{V}+\kappa^{*}q+\kappa q^{*} \).
The small term \( (\kappa^{*}q+\kappa q^{*}) \) matters in the case \
\( \lambda _{i}=\lambda _{f}=\pm 1 \)
only. 
For illustration it is enough to do one case, 
\( \gamma ^{*}V,\lambda _{i}=0,\lambda _{f}=1 \),
explicitely: \( W_{1}=0,W_{0}=-yQ\kappa, \)
\be
\Phi ^{V,01}|_{z_{0}}=\int ^{z_{0}}_{0}z\overline{z}
\varphi ^{V,01}_{4}(\kappa,z,q)dz\approx 
 \pi \frac{m^{2}_{V}Q}{(m^{2}_{_{^{V}}}+Q^{2})} 
\ee
$$
\left\{ \frac{\kappa |\kappa|^{2}}{Q^{4}}
\ln\frac{Q^{2}z_{0}+|\kappa|^{2}}{|\kappa|^{2}}-
\frac{(\kappa+q)|\kappa+q|^{2}}{Q^{4}}
\ln\frac{Q^{2}z_{0}+|\kappa+q|^{2}}{|\kappa +q|^{2}}-
\frac{q|q|^{2}}{Q^{4}}
\ln\frac{Q^{2}z_{0}+|q|^{2}}{|q|^{2}}\right\}  
$$

In the kinematics \( z_{0}Q^{2}\gg |\kappa|^{2}\gg |q|^{2} \) this can
be approximated by 
\be
\Phi ^{V,01}|_{z_{0}}\approx 
\pi \frac{m^{2}_{V}Q}{(m^{2}_{_{^{V}}}+
Q^{2})^{3/2}}\left\{ \frac{f^{(3)}(\kappa,q)}{(m_{V}^{2}+Q^{2})^{3/2}}
\ \ \ln\frac{Q^{2}z_{0}}{|\kappa|^{2}}\right\} 
\ee

In this way we obtain (\ref{end-point}) confirming the matching with
the expression (\ref{main}) for \( z={\cal O}(1) \), 
i.e. the cancellation of the auxiliary
\( z_{0} \).

\end{document}